\documentstyle[12pt]{article}
\font\fontone= cmr10 scaled 2000

\font\fontr=msbm10 scaled 1000
\def \dst {\displaystyle}

\font\gotic=eufm10 scaled 1100

\newcommand{\la} {\lambda}
\newcommand{\lt} {\ {\bf <}}
\newcommand{\rt} {{\bf >}\ }
\newcommand{\lb} {\left(}
\newcommand{\rb} {\right)}
\newcommand{\bl} {\Bigl(}
\newcommand{\br} {\Bigr)}
\newcommand{\ld} {\left.}
\newcommand{\rd} {\right.}

\newcommand{\al} {\alpha}

\newcommand{\be} {\beta}

\newcommand{\ga} {\gamma}
\newcommand{\bga}{\begin{array}{l}}
\newcommand{\ena}{\end{array}}
\newcommand{\bge}{\begin{equation}}
\newcommand{\ene}{\end{equation}}

\author{M. Zyskin \thanks{BRIMS, Hewlett Packard Labs, Filton Road, Stoke
Gifford, Bristol, BS34 8QZ, UK;   maxim@hplb.hpl.hp.com}}
\title{On gauge fields - strings duality as an integrable system.}
\date{}
\begin{document}
\maketitle
\begin{abstract}
It was suggested in \cite{hep-th/0002106}, that semiclassically, a partition
function of a string theory in the 5 dimensional constant negative curvature
space with a boundary condition at the absolute satisfy the loop equation with
respect to varying the boundary condition, and thus the partition function of the string 
gives the expectation value of a Wilson loop in the 4 dimensional QCD. In the paper, 
we present the geometrical framework, which reveals that the equations of motion of the string 
are integrable, in the sence that they can be written via a Lax pair with a spectral parameter.
We also show, that the issue of the loop equation rests solely on the properly posing the
boundary condition. 
\end{abstract}
\newpage

\section{Introduction and summary.}
\newcounter{prop}
\setcounter{prop}{0}
\renewcommand{\theprop}{\thesection.\arabic{prop}}

According to \cite{hep-th/0002106}, \cite{p2}, there is a semiclassical
evidence, that
the expectation value of a QCD Wilson loop in dimension 4 is given by a
partition function of a string theory
with the 5 dimensional constant negative curvature target space and with a
boundary condition on the loop at the absolute,
as the  partition function satisfy the loop equation with respect to variation
of the boundary loop.
Semiclassically, the equations of motion are the equations of the minimal
surface in the constant negative curvature 
space with prescribed boundary; and the action is the area of such minimal
surface (we sum over minimal surfaces when the
solution is not unique). In the paper, we present the equations of motion for
such string theory in a  geometrical
format, which reveals, that those equations are integrable,
since they can be written via a Lax pair
with a spectral parameter, and therefore, the methods of boundary integrable
models might be applicable, either modern ones
like boundary inverse scattering \cite{ff}, \cite{fz}; boundary r-matrices, the bootstrap,
etc  \cite{z};  
or rather  more geometrical
ones, as they used more then 100 years
ago to solve nonlinear partial 
differential equations such as Liouville, Monge Ampere, and sine Gordon. 
We also show that the issue of the loop equation being satisfied rests solely on the properly posing the
boundary condition, since the terms in the second variation containing $\delta$ function 
is given by  a line integral over the boundary.

Some other  geometric and integrability aspects of the methods used in the paper 
will be published 
elswhere, \cite{inworks}, \cite{fz}. 

\section{Moving Frames; surfaces in $H_{(n)}$; minimal surfaces.}
\subsection{Frames in $M^{n+1}$}
Let $M^{n+1}$ be Minksowsky space: a real vector space with the metric $\eta =
(-1,1,1,\ldots 1)$.
A {\it frame} in $M^{n+1}$ is an $(n+1)$-tuple of vectors 
\bge
\bga
F= (f_{(0)}, f_{(1)},f_{(2)},\ldots f_{(n)} ) , \ \mbox{such that }\\
\lt f_{(0)},f_{(0)}\rt = -1, \  \lt f_{(\al )},f_{(\al )}\rt = 1, \ \al = 1,2,
\ldots n.
\label{sp}
\ena
\ene
A standart frame is the following  $(n+1)$ tuple of vectors:
$$
\bga
e_{(0)} = (1,0,0,\ldots 0 )\\
e_{(1)} = (0,1,0,\ldots 0 )\\
\ldots \\
e_{(n)} = (0,0,0,\ldots 1 )
\ena
$$
Any frame is obtained from the standart frame by an action of the group
$SO(1,n)$, $f_{(i)} = g_{ij} e_{(j)}$, or in components,
\bge\pmatrix{
f_{(0)}^{0}  & f_{(0)}^{1} & f_{(0)}^{2} & \ldots & f_{(0)}^{n} \cr
f_{(1)}^{0}  & f_{(1)}^{1} & f_{(1)}^{2} & \ldots & f_{(1)}^{n} \cr
\ldots       & &&& \cr
f_{(n)}^{0}  & f_{(n)}^{1} & f_{(n)}^{2} & \ldots & f_{(n)}^{n} \cr
} \ = \ \mbox{{\fontone G}} \ 
\pmatrix{
1 & 0 & 0 & \ldots & 0 \cr
0 & 1 & 0 & \ldots & 0 \cr
\ldots       & &&& \cr
0 & 0 & 0 & \ldots & 1 \cr
},
\label{g}
\ene
since (~\ref{g}) defines a matrix  {\bf G} with  $  {\bf{{\fontone G}} \ {\bf
\eta} \  \bf{{\fontone G}} }^{T} = \bf{ \eta}$, $\eta 
= Diag \ (-1,1,1,\ldots 1)$, and therefore, $ {\bf G} \in \mbox{ SO }(1,n) $.

\subsection{The hyperboloid $H_{(n)}$}
Since $\lt f_{(0)},f_{(0)} \rt \equiv  - {f_{(0)}^0}^2 + {f_{(0)}^1}^2 + \ldots
 {f_{(0)}^n}^2
 = -1$, the vector $f_{(0)}$ can be identified with a point on a hyperboloid
$H_{(n)}$ in $M^{n+1}$, 
\bge
- {(x_0 )}^2 + {(x_1  )}^2 + {(x_2)}^2 + \ldots {(x_n )}^2 = - 1
\label{H}
\ene
 and for a fixed $f_{(0)}$, the tangent space to the hyperboloid at the point
$f_{(0)}$ 
is spanned by $\{ f_{(1)}, f_{(2)}, \ldots f_{(n)} \}$; indeed, for any $\al
=1,2,\ldots n$, 
$$
\lt f_{(0)} + \epsilon f_{(\al)}, f_{(0)} + \epsilon f_{(\al)} \rt =  \lt
f_{(0)},f_{(0)} \rt  + 2 \epsilon 
\lt f_{(0)}, f_{(\al)}\rt + \epsilon^2 \lt f_{(\al )}, f_{(\al)}\rt = 
-1 + \epsilon^2,
$$
as $\lt f_{(0)}, f_{(\al)}\rt =0 $.

\vspace{7mm}

Suppose at each point of a hyperboloid a frame $F$ is choosen, with $f_{(0)}$
corresponding to the point itself,
and other $f_{(\al)}$ choosen arbitrary in the tangent space to the
hyperboloid; obviuosly it is always possible to
do. Since $ {\bf f } = {\bf G} \ {\bf e}$, ( which is components is $ { f_{(i)}
} = \dst\sum_{j} { g}_{ij} \ { e_{(j)}}$ ),
$$
d {\bf f} = d {\bf G} \ \ {\bf e} =  \left( d {\bf G }\ {\bf G}^{-1} \right)  \
{\bf f} \equiv {\bf \omega} \ {\bf f},
$$
or $ d f_{(i)} = \omega_{ij} f_{(j)}$, where $\omega_{ij} = \left( d {\bf G }\
{\bf G}^{-1} \right)_{ij}$; 
therefore, \\
$ d \omega = d \left( d {\bf G }\ {\bf G}^{-1} \right) = d {\bf G }\ {\bf
G}^{-1} \wedge d {\bf G }\  {\bf G}^{-1} = 
\omega \wedge \omega $, \\
and we arrived at the Maurer-Cartan equations:
\bge
\bga
d {\bf f} = {\bf \omega} \ {\bf f} 
\label{fr}
\ena
\ene
\bge
\bga
d \omega = \omega \wedge \omega
\label{mc}
\ena
\ene
In fact, if (~\ref{mc}) is satisfied, then equations (~\ref{fr}) 
are comptaible. It's easy to see that
(~\ref{mc}) is necessary, as if $d {\bf f} = {\bf \omega} \ {\bf f} $, 
it follows that 
$ 0 = d ({\bf \omega} \ {\bf f})= (d \omega  \omega \wedge \omega )  \ {\bf f})$.
Therefore, the crucial part in what follows will be to construct one forms satisfying 
(~\ref{mc}), as then the frame can be found by integrating the compatible first order equations (~\ref{fr}).

From (~\ref{sp}), it follows that
\bge
\bga
\omega_{0, 0} = 0\\
\omega_{0, \al} = \omega_{\al, 0} \\
\omega_{\al, \be} = - \omega_{\be, \al}, \quad \al, \be =1,2,\ldots n.\\
\ena
\ene 
\vspace{7mm}

The  induced metric on the hyperboloid is just
\bge
h:=\lt d f_{(0)}, \otimes d f_{(0)} \rt = \dst\sum \omega_{0,i} \otimes
\omega_{0,j} \lt f_{(i)}, f_{(j)} = 
\dst\sum {\omega_{0,i}}^{\otimes 2}
\ene

Since the curvature 2-form on the hyperboloid is
$$- \Omega_{\al, \be} = d \omega_{\al, \be} - \omega_{\al, \ga}\wedge
\omega_{\ga, \be}, \quad \al, \be, \ga =1,2,\ldots n,$$ and from Maurer Cartan
(~\ref{mc}) it follows that
$$\Omega_{\al, \be}= - \omega_{0,\al} \wedge \omega_{0,\be}$$
It is easy to see that the hyperboloid has constant negative curvature in this
language: 
choose the basis of tangent vectors $X_i$, $i =1,2,\ldots$, such that $
\omega_{0, i } ( X_j) = \delta_{ij}$.
Then such vectors are ortonormal in the induced metric, $h (X_i, X_j) =
\delta_{ij}$. The Riemann tensor in this basis is $R_{ijkl}= \Omega_{ij} (X_k,
X_l)$.
Contracting two indices, \\
$R_{kikj}= \Omega_{ki} (X_k, X_j) = - \omega_{0,k} \wedge \omega_{0,i} (X_k,
X_j)= - (n-1) \delta_{i,j} = -(n-1) h (X_i, X_j)$.

\subsection{A surface in $H_{(n)}$}
Since $f_{(0)}$ is identified with a point on a hyperboloid, 
a vector- valued function of 2 real varibles $f_{(0)}\  (u,v) $ defines  a
surface on the hyperboloid. 
At each point on the surface, we  choose the frame $F(u,v)$ , (~\ref{sp}), in
such a way, that $f_{(1)}$ and $f_{(2)}$ 
will span the tangent space of the surface (of course, there are many ways to
do it); as before, all  $\{ f_{(i)}(u,v)\}, \quad i=1,2,\ldots n$ span the
tangent space at $f_{(0)}$,  $T_{f_{(0)}}$ in the hyperboloid $H_{(n)}$. 
With this choise,
\bge
d f_{(0)}  (u,v) = \lb  \omega_{01} f_{(1)}  + \omega_{02} f_{(2)} \rb (u,v)
\label{df0}
\ene
and 
\bge
\omega_{0\mu } =0, \mu =3,4,\ldots n ;
\ene
{}From Maurer-Cartan also 
$d \omega_{0\mu } = \omega_{01} \wedge \omega_{1\mu} + \omega_{02} \wedge
\omega_{2 \mu} = 0 $
{}From this follows that 
\bge
\bga
\omega_{1\mu} = b_{(\mu), 1} \omega_{01} + c_{(\mu)} \omega_{02}\\
\omega_{2\mu} = c_{(\mu)} \omega_{01} + b_{(\mu), 2} \omega_{02},
\label{bc}
\ena
\ene
where $ b_{(\mu), \al}$ and $c_{(\mu)}$ are some functions on the surface. 

The first fundamental form on the surface is
\bge
I = \lt d f_{(0)}, \otimes d f_{(0)} \rt =  \omega_{01}^{\otimes 2} +
\omega_{02}^{\otimes 2}
\ene

To each normal (in $M^{n+1} $)direction $\mu = 3,4,\ldots$ there correspond a
second fundamental form,
\bge
II_\mu = \omega_{1\mu} \otimes \omega_{01} + \omega_{2\mu} \otimes \omega_{02}
\label{2ff}
\ene

It's convinient to introduce notations
\bge
\omega_{ij} = \al_{ij} du + \be_{ij} du
\label{albe}
\ene

We choose the conformal coordinates on the surfaces, such that
\bge
I = e^{2 \phi} \lb du^2 + dv^2 \rb
\label{cfg}
\ene
In this coordinates, $\lt  \frac{\partial}{\partial u } f_{(0)},
\frac{\partial}{\partial v } f_{(0)}\rt =0$, and therefore we can choose 
$f_{(1)}$ and  $f_{(2)}$ in such a way that 
$$
d f_{(0)}  (u,v) = \lb \omega_{01} f_{(1)}  + \omega_{02} f_{(2)} \rb (u,v),
$$
with $\al_{02}=0, \quad \be_{01}=0$ in notations (~\ref{albe}).  In conformal
coordinates (~\ref{cfg}), we also have
${(\al_{01})}^2 = e^{2 \phi}$ and ${(\be_{02})}^2 = e^{2 \phi}$, thus we can make a choise
\bge
\omega_{01}  = e^{ \phi} du ; \quad \omega_{02}  = e^{ \phi} dv
\label{0102}
\ene
{}From (~\ref{mc}), (~\ref{df0}) we have 
$$
\bga
d \omega_{01}  = - \omega_{02} \wedge \omega_{12} \\
d \omega_{02}  =  \omega_{01} \wedge \omega_{12},
\ena
$$
and from (~\ref{albe}), (~\ref{0102}) it follows
\bge
\begin{array}{rl}
\al_{01} = e^{ \phi}, & \be_{01} =0 \cr
\al_{02} = 0 , &  \be_{02} = e^{ \phi}\cr
\al_{12} = - \frac{\partial \phi}{\partial v } , & \be_{12} =  \frac{\partial
\phi}{\partial u }  \cr
\al_{1\mu} = b_{(\mu), 1} e^{ \phi}, & \be_{1\mu} =  c_{(\mu)} e^{ \phi} \cr
\al_{2\mu} =  c_{(\mu)} e^{ \phi} , &  \be_{2\mu} = b_{(\mu), 2} e^{ \phi}\cr
\ena
\label{010212}
\ene
{}From (~\ref{mc}), (~\ref{df0}) 
$$
d \omega_{12}  = \omega_{01} \wedge \omega_{02} -\dst\sum_{\mu = 3, 4, \ldots}
\omega_{1\mu} \wedge \omega_{2\mu},
$$
and therefore, using (~\ref{bc}), 

\bge
\lb \frac{\partial^2}{\partial u^2 } + \frac{\partial^2}{\partial v^2 } \rb \
\phi =
e^{2 \phi} \lb 1 + \dst\sum_{\mu = 3, 4, \ldots} \ det 
\pmatrix{b_{(\mu), 1} & c_{(\mu)}  \cr
c_{(\mu)} & b_{(\mu), 2} \cr} .
\rb
\ene

\bge
\bga
d \omega_{1\mu}  = \omega_{12} \wedge \omega_{2\mu} +
\dst\sum_{\nu = 3, 4, \ldots} \omega_{1\nu} \wedge \omega_{\nu\mu}\\
d \omega_{2\mu}  = - \omega_{12} \wedge \omega_{1\mu} +
\dst\sum_{\nu = 3, 4, \ldots} \omega_{2\nu} \wedge \omega_{\nu\mu}
\ena
\ene

\bge
\bga
2 c_{(\mu)} \frac{\partial \phi}{\partial u } - \lb b_{(\mu), 1} - b_{(\mu), 2}
\rb \frac{\partial \phi}{\partial v } +
 \frac{\partial c_{(\mu)} }{\partial u } \phi -  \frac{\partial b_{(\mu), 1}
}{\partial v }\phi =
\dst\sum_{\nu = 3,4,\ldots} \lb b_{(\nu), 1} \be_{\nu \mu}- c_{(\nu)} \al_{\nu
\mu}\rb 
\\
\lb b_{(\mu), 2} - b_{(\mu), 1} \rb  \frac{\partial \phi}{\partial u } - 2
c_{(\mu)} \frac{\partial \phi}{\partial v } +
 \frac{\partial b_{(\mu), 2} }{\partial u } \phi -  \frac{\partial c_{(\mu)}
}{\partial v }\phi =
\dst\sum_{\nu = 3,4,\ldots} \lb c_{(\nu)}  \be_{\nu \mu}-  b_{(\nu), 2}
\al_{\nu \mu}\rb
\\
\mu = 3,4,\ldots
\ena
\ene

\bge
d \omega_{\mu \nu}  = - \omega_{1 \mu } \wedge \omega_{1 \nu} - \omega_{2 \mu }
\wedge \omega_{2 \nu}
\dst\sum_{\eta = 3, 4, \ldots} \omega_{\mu \eta } \wedge \omega_{\eta \nu}\\
\ene

\bge
\frac{\partial \be_{\mu \nu} }{\partial u } - \frac{\partial \al_{\mu \nu}
}{\partial v }
=
- e^{ 2 \phi} 
\bl ( b_{(\mu), 1} - b_{(\mu), 2} ) c_{(\nu)}  - ( b_{(\nu), 1} - b_{(\nu), 2}
)  c_{(\mu)} \br +
\dst\sum_{\eta = 3, 4, \ldots} ( \al_{\mu \eta }  \be_{\eta \nu} - \be_{\mu
\eta }  \al_{\eta \nu} )
\ene

\subsection{Lagrangians, Variational derivative, minimal Surface}
The equations of motion  are just the equations for the minimal surface. They
can be obtained from the condition that the variation of the Lagrangian
\bge
L = \dst\int \sqrt{ Det_{\al \be} \lb  \lt \dst\frac{\partial}{\partial u_\al}
f_{(0)}, \dst\frac{\partial}{\partial u_\be} f_{(0)} \rt   \rb } du_1 \wedge
du_2;
\ene
is zero, $\dst\frac{ \delta L}{\delta f_{(0)}} = 0 $, 
subject to $\lt \delta f_{(0)}, f_{(0)}\rt =0 $; the last ensures that we stay
on the hyperboloid. In the conformal coordinates (~\ref{cfg}), those equations
are
$$
\bga
0 = \lt \delta f_{(0)}, \dst\frac{\partial}{\partial u} \dst\frac{\partial
f_{(0)} }{\partial u} + \dst\frac{\partial}{\partial v} \dst\frac{\partial
f_{(0)} }{\partial v}\rt = 
\\
\lt \delta f_{(0)}, \dst\frac{\partial}{\partial u} \lb e^\phi f_{(1)} \rb +
\dst\frac{\partial}{\partial v} \lb e^\phi f_{(2)} \rb \rt =\lt \delta f_{(0)},
e^{2 \phi} \lb  \# f_{(0)} + \dst\sum_{\mu =3,4,\ldots } ( b_{(\mu) 1} +
b_{(\mu)2} )
 f_{(\mu)} \rb \rt;
\ena
$$
we made the computation in the conformal  basis, (~\ref{0102}), and used the Maurer Cartan equations, 
(~\ref{010212}). 
Since $\lt \delta f_{(0)}, f_{(0)}\rt =0 $, and otherwise arbitrary, it follows
that
\bge
( b_{(\mu) 1} + b_{(\mu)2} ) = 0, \mu = 3,4,\ldots
\label{meancurvat}
\ene

\subsection{Minimal surface in $H_{(3)}$ as an integrable system}

The Maurer-Cartan equations for the minimal surface (~\ref{meancurvat})
simplify, and for the surface
in $H_{(3)}$ they are
\bge
\bga
\lb \frac{\partial^2}{\partial u^2 } + \frac{\partial^2}{\partial v^2 } \rb \
\phi =
e^{2 \phi} \lb 1 + b^2 +c^2 \rb, 
\label{coshg}
\ena
\ene

\bge
\bga
2 c \phi_u - 2 b \phi_v + c_u - b_v = 0 , \\
2 b \phi_u + 2 c \phi_v + b_u + c_v = 0,
\label{h3}
\ena
\ene
where $b\equiv b_{(3) 1} = - b_{(3) 2} , c \equiv c_{(3)}$; and $\al_{12},
\be_{12}$ are determined by $\phi$,
$\al_{12} = - \phi_v, \be_{12} = \phi_u $. The system (~\ref{h3}) is
integrable; it has a Lax pair,  with a spectral parameter $\la \in \mbox{\fontr C}$, for example 
this one (there is in fact a much better one for purposes of inverse scattering; but the one below
is more geometric):

\bge
\bga
\dst\frac{\partial}{\partial u}  \Phi  = 
\left[ \matrix{
0 & \frac{\la^2 + 1 }{2 \la }    e^\phi & \frac{- i  \la^2 + i  }{2 \la }
e^\phi & 0 \cr
\frac{\la^2 + 1 }{2 \la }    e^\phi  & 0 & - \phi_v & \frac{\la^2 (b- i c ) +
(b+ i c)  }{2 \ \la } e^{\phi} 
 \cr
\frac{- i  \la^2 + i  }{2 \la }    e^\phi & \phi_v & 0 & \frac{ \la^2 (c + i b
) + (c - i b)   }{2 \la } e^{\phi} 
\cr
0 & - \frac{\la^2 (b- i c ) +  (b+ i c)  }{2 \ \la } e^{\phi}  & - \frac{ \la^2
(c + i b ) + (c - i b)   }{2 \la } e^{\phi} & 0 \cr}
\right]  \Phi 
\\[20mm]
\dst\frac{\partial}{\partial v}  \Phi  = 
\left[ \matrix{
0 &  \frac{i  \la^2 - i }{2 \la }    e^\phi & \frac{\la^2 + 1 }{2 \la }
e^\phi  & 0 \cr
\frac{i  \la^2 - i }{2 \la }    e^\phi  & 0 &  \phi_u & \frac{\la^2 (c + i b )
+  (c - i b )  }{2 \ \la } e^{\phi} 
 \cr
\frac{\la^2 + 1 }{2 \la }    e^\phi & - \phi_u & 0 & \frac{ \la^2 (- b + i c )
+ (-b - i c)   }{2 \la } e^{\phi} 
\cr
0 & - \frac{\la^2 (c + i b ) +  (c - i b )  }{2 \ \la } e^{\phi} 
 & - \frac{ \la^2 (- b + i c ) + (-b - i c)   }{2 \la } e^{\phi} & 0 \cr}
\right]  \Phi 
\ena
\ene
	In fact, the integrable system here is something quite familiar. It follows from (\ref{h3}),
that 
\bge\bga
\phi_u = - \dst\frac{ b b_u + c c_u + b c_v -c b_v}{2 (b^2 + c^2)} \\
\phi_v = - \dst\frac{ b b_v + c c_v - b c_u + c b_u}{2 (b^2 + c^2)};\label{phi3}
\ena
\ene
here subscripts $u$ and $v$ denote derivatives with respect to $u, v$.
Let's introduce $\rho$ and $\Theta$, such that $ b = \rho \cos \Theta$, $ c = \rho \sin \Theta$.
Then it follows from (~\ref{phi3} ) that
$$
\bga
{\lb  \phi + \frac{1}{2} \log \rho \rb}_u = - \Theta_v, \\
{\lb  \phi + \frac{1}{2} \log \rho \rb}_v =  \Theta_u.
\ena
$$
Therefore, $\Theta$ must be a harmonic function of $(u,v)$, 
( as well as $\lb  \phi + \frac{1}{2} \log \rho \rb$), 
so whenever the only harmonic functions are  constants;
say if a surface is an imbedding of a sphere; then  
$\lb  \phi + \frac{1}{2} \log \rho \rb$ is some  constant $\kappa$ as well;
and so the equation (~\ref{coshg}) is in fact a $cosh$ -Gordon,
\bge
\lb \frac{\partial^2}{\partial u^2 } + \frac{\partial^2}{\partial v^2 } \rb \
\phi =
e^{2 \phi} + \kappa e^{-2 \phi}.
\ene
\subsection{A remark on minimal surfaces in $\mbox{\fontr R}^3$ and the Liouville equation.}
It is well known that a minimal surface in $\mbox{\fontr R}^3$ is a 
surface with the  mean curvature equal to zero. In the Maurer Cartan format, the equations for
a surface in $\mbox{\fontr R}^3$ are
\bge
\bga
dx = \omega_i  f_{(i)}\\
d f_{(i)} = \omega_{ij} f_{(j)} \\[7mm]
d \omega_i = \omega_j \wedge \omega_{ji}\\
d \omega_{ij} = \omega_{ik} \wedge \omega_{kj},
\ena
\ene
where $x\in \mbox{\fontr R}^3$, and $\{f_{(\mu)}\}$ is an ortonormal frame; the group is the group 
of Eucledean motions of $\mbox{\fontr R}^3$, instead of Lorents group which we work with.
For a minimal surface,
choosing the conformal coordinates (~\ref{cfg}), writing the Maurer-Cartan equations, and taking 
into account that the mean curvature is zero, similar to what we did in constant negative curvature
space above, we would arrive at the Liouville equation, 
$$
\lb \frac{\partial^2}{\partial u^2 } + \frac{\partial^2}{\partial v^2 } \rb \
\phi =
e^{- 2 \phi}, 
$$ 
for which a solution can be written explicitly,
as it is well known for a very long time; but the corresponding quantum field theory is regarded to be 
notoriously difficult \cite{z}. 

I find quite amusing the following set of facts: a) 
Maurer Cartan plays a major role
in the geometry of frames on surfaces, and in particular it is
 responsible for the Liouville equation; b) some Maurer Cartan shows up  in the
 celebrated deformation quantization construction 
of associative algebras,  c) the way they approach quantum Liouville in \cite{z} is via associativity
of the operator product algebra, and d) they seem to be using the same software to draw their pictures 
in their texts in b and c, and if you look at those pictures from far away, they look alike; 
but I do not know what exactly to make of those observations. 

For a surface of constant mean curvature h, we would obtain the
 $sinh$ Gordon, $\lb \frac{\partial^2}{\partial u^2 } + \frac{\partial^2}{\partial v^2 } \rb \
\phi =
- h^2 e^{2 \phi} + e^{-2 \phi})$. This and other integrable surfaces in $\mbox{\fontr R}^3$ 
were studied in \cite{gf1}.

\subsection{Minimal surface in $H_{(5)} $ as an integrable system}
The Maurer-Cartan equations for the minimal surface (~\ref{meancurvat}) 
in $H_{(5)}$  are
\bge
\bga
\lb \frac{\partial^2}{\partial u^2 } + \frac{\partial^2}{\partial v^2 } \rb \
\phi =
e^{2 \phi} \lb 1 + {b_3}^2 + {c_3}^2 + {b_4}^2 + {c_4}^2   +  {b_5}^2 +
{c_5}^2\rb,  \\[7mm]
L_{1\mu} \stackrel{(def.)}{=} 2 c_{(\mu)} \frac{\partial \phi}{\partial u } - 2
 b_{(\mu)} \frac{\partial \phi}{\partial v } +
 \frac{\partial c_{(\mu)} }{\partial u } \phi -  \frac{\partial b_{(\mu)}
}{\partial v }\phi =
\dst\sum_{\nu = 3,4,5, \ \nu\neq \mu } \lb b_{(\nu)} \be_{\nu \mu}- c_{(\nu)}
\al_{\nu \mu}\rb 
\\
L_{2\mu} \stackrel{(def.)}{=}  - 2 b_{(\mu)}  \frac{\partial \phi}{\partial u }
- 2 c_{(\mu)} \frac{\partial \phi}{\partial v } -
 \frac{\partial b_{(\mu)} }{\partial u } \phi -  \frac{\partial c_{(\mu)}
}{\partial v }\phi =
\dst\sum_{\nu = 3,4,5 \ \nu\neq \mu } \lb c_{(\nu)}  \be_{\nu \mu} +  b_{(\nu)}
\al_{\nu \mu}\rb, 
\mu = 3,4,5;
\label{h5}
\ena
\ene
and 

\bge
\bga
\frac{\partial \be_{\mu \nu} }{\partial u } - \frac{\partial \al_{\mu \nu}
}{\partial v }
=
- 2  e^{ 2 \phi} 
\bl   b_{(\mu)} c_{(\nu)}  -    b_{(\nu)}  c_{(\mu)} \br +
\dst\sum_{\eta = 3,4,5 \ \eta \neq \mu, \nu} ( \al_{\mu \eta }  \be_{\eta \nu}
- \be_{\mu \eta }  \al_{\eta \nu} )\\
\mu , \nu = 3,4,5.
\ena
\ene

This system of equations appear integrable, and posess a Lax pair with spectral
parameter, as follows.
We assume that  say $\al_{45}, \be_{45}$ can be
represented in the form
$$
\al_{45} = \psi_u  + \chi_v , \quad \be_{45} = \psi_v  - \chi_u
$$
with certain functions $\psi(u,v), \chi(u,v)$; which doesnot seem to be terrribly restrictive.
 There is a  Lax pair, reproducing the Maurer Cartan equations; it  involves a
spectral parameter
$\la \in \mbox{fontr C}$, and the  unknowns: $\psi$, $\chi$,
the conformal factor  $\phi(u,v)$, as well as 
$\{c_m (u,v), b_m (u,v)\vert m=3,4,5 \}$, see (~\ref{bc}), 
 (~\ref{2ff}), (where $b_m (u,v) \stackrel{(def)}{=}  b_{(m),1} (u,v)  = -
b_{(m),2} (u,v)$, as the surface is minimal); that's  all we need to know to 
be able to find  the Maurer-Cartan 1-forms,
(~\ref{mc}), and then the surface itself is obtained by solving linear compatible  first 
order equations (~\ref{fr}). Possibly, there are better, for purposes of boundary inverse 
scattering, Lax pairs; this is under investigation; but at least, there is some Lax pair:

\bge
\bga
\dst\frac{\partial}{\partial u}  \Phi  = \\ [5mm]
\left[ \matrix{
0 & \frac{\la^2 + 1 }{2 \la }    e^\phi & \frac{- i  \la^2 + i  }{2 \la }
e^\phi  &  & ( 0 \  0 \  0) \cr
\frac{\la^2 + 1 }{2 \la }    e^\phi & 0 &  - \phi_v & & \lb \left. {\frac{\la^2
(b_\mu- i c_\mu ) +  (b_\mu+ i c_\mu)  }{2 \ \la } e^{\phi}}
\right|_{\mu=3,4,5} \rb  \cr
\frac{- i  \la^2 + i  }{2 \la }    e^\phi& \phi_v & 0 & & \lb \ld \frac{ \la^2
(c_\mu + i b_\mu ) + (c_\mu - i b_\mu)   }{2 \la } e^{\phi}\right|_{\mu=3,4,5}
\rb  
  \cr
0 & - \frac{\la^2 (b_3- i c_3 ) +  (b_3+ i c_3)  }{2 \ \la } e^{\phi}  & -
\frac{ \la^2 (c_3 + i b_3 ) + (c_3 - i b_3)   }{2 \la } e^{\phi} &  &  \cr
0 & - \frac{\la^2 (b_4- i c_4 ) +  (b_4+ i c_4)  }{2 \ \la } e^{\phi}  & -
\frac{ \la^2 (c_4 + i b_4 ) + (c_4 - i b_4)   }{2 \la } e^{\phi} &  &
\mbox{\fontone A }  \cr
0 & - \frac{\la^2 (b_5- i c_5 ) +  (b_5+ i c_5)  }{2 \ \la } e^{\phi}  & -
\frac{ \la^2 (c_5 + i b_5 ) + (c_5 - i b_5)   }{2 \la } e^{\phi} &  &  \cr
} \right]  
 \Phi 
\\[30mm]
\mbox{\fontone A } = \left[ \matrix{
0 & a[3,4]  &  a[3,5] \cr 
- a[3,4] & 0  & ( \psi_u  + \chi_v ) \cr 
- a[3,5] & - ( \psi_u  + \chi_v )  &  0 \cr 
}
\right] \\[15mm]
a[3,4] = \frac{  1 }{
({ b_3 }^2 + { c_3 }^2) } 
\lb - c_3
 \bl L_{14} -c_5 (\psi_u  + \chi_v ) +b_5 (\psi_v  - \chi_u) \br + \rd
\\
 \ld b_3 \bl  L_{24} +b_5 (\psi_u  + \chi_v ) +c_5 (\psi_v  - \chi_u) \br \rb
\\[10mm]
a[3,5] = \frac{1}{
({ b_5 }^2 + { c_5 }^2)} \bl 
  { c_5 } L_{13} - { b_5 } L_{23}\br +  \\[3mm]
\frac{1}{
({ b_3 }^2 + { c_3 }^2) ({ b_5 }^2 + { c_5 }^2)} \bl \lb { c_3 } ( { b_4 } {
b_5 } + { c_4 } { c_5 }) + 
   { b_3 } ( { b_4 } { c_5 } - { b_5 } { c_4 }) \rb ( L_{14} -c_5 (\psi_u  +
\chi_v ) +b_5 (\psi_v  - \chi_u) ) \\
    + \lb  { c_3 } ( { b_4 }{ c_5 } -  
   { b_5 } { c_4 }) -
   { b_3 }( { c_4 } { c_5 } + { b_4 } { b_5 })\rb ( L_{24} +b_5 (\psi_u  +
\chi_v ) +c_5 (\psi_v  - \chi_u) )\br 
\ena
\ene
\vspace{10mm}

\bge
\bga
\dst\frac{\partial}{\partial v}  \Phi  = \\[6mm]
\left[ \matrix{
0 & \frac{i  \la^2 - i }{2 \la }    e^\phi & \frac{\la^2 + 1 }{2 \la }
e^\phi  &  & ( 0\  0 \  0) \cr
\frac{i  \la^2 - i }{2 \la }    e^\phi  & 0 &  \phi_u & & \lb \frac{\la^2 (
c_\mu + i b_\mu) +  (c_\mu - i b_\mu)  }{2 \ \la } e^{\phi} \rb 
 \cr
\frac{\la^2 + 1 }{2 \la }    e^\phi& - \phi_u & 0 & & \lb \frac{ \la^2 ( -
b_\mu + i c_\mu ) + (-  b_\mu - i c_\mu )   }{2 \la } e^{\phi} \rb  
\cr
0 & - \frac{\la^2 (c_3 + i b_3  ) +  (c_3 - i b_3 )  }{2 \ \la } e^{\phi}  & -
\frac{ \la^2 ( -b_3 + i c_3 ) + 
(-b_3 - i c_3 )   }{2 \la } e^{\phi} &  &  \cr
0 & - \frac{\la^2 (c_4 + i b_4  ) +  (c_4 - i b_4 )  }{2 \ \la } e^{\phi}  & -
\frac{ \la^2 ( -b_4 + i c_4 ) + 
(-b_4 - i c_4 )   }{2 \la } e^{\phi} &  & \mbox{\fontone B }  \cr
0 & - \frac{\la^2 (c_5 + i b_5  ) +  (c_5 - i b_5 )  }{2 \ \la } e^{\phi}  & -
\frac{ \la^2 ( -b_5 + i c_5 ) + 
(-b_5 - i c_5 )   }{2 \la } e^{\phi} &  &  \cr
}
\right]  \Phi 
\\[30mm]
\mbox{\fontone B } = 
\left[ \matrix{
0 & b[3,4]  &  b[3,5] \cr 
- b[3,4] & 0  & ( \psi_v  - \chi_u ) \cr 
- b[3,5] & - ( \psi_v  - \chi_u )  &  0 \cr 
}
\right] \\[10mm]
b[3,4] = \frac{1}{
({ b_3 }^2 + { c_3 }^2) } \bl 
{ b_3 } ( L_{14} -c_5 (\psi_u  + \chi_v ) +b_5 (\psi_v  - \chi_u) ) +  \\
  { c_3 }  ( L_{24} +b_5 (\psi_u  + \chi_v ) +c_5 (\psi_v  - \chi_u) )\br
\\[10mm]
b[3,5] = \frac{1}{
({ b_5 }^2 + { c_5 }^2) } \bl 
-{ b_5 }   L_{13}   - { c_5 } L_{23}\br  + \\
\frac{1}{
({ b_3 }^2 + { c_3 }^2) ({ b_5 }^2 + { c_5 }^2)} \bl 
\lb  
   { b_5 } ( { c_3 } { c_4 } - { b_3 } { b_4 } ) + 
    { c_5 } ( { b_4 } { c_3 } - { b_3 } { c_4 } )\rb ( L_{14} -c_5 (\psi_u  +
\chi_v ) +b_5 (\psi_v  - \chi_u) ) \\
+ 
   \lb  { b_5 } ( { b_3 } { c_4 } -{ b_4 } { c_3 })  - 
  { c_5 }  ( { c_3 } { c_4 }  + { b_3 } { b_4 })  \rb  ( L_{24} +b_5 (\psi_u  +
\chi_v ) +c_5 (\psi_v  - \chi_u) )\br
\ena
\ene
where $L_{ij}$ are defined in (~\ref{h5}).

\section{Some thoughts on the loop equation, in the context of zero mean curvature surfaces.}
We currently do not know how to pose an inverse scattering problem for the equations we got; 
however, an experience with an inverse scattering on an arbitrary domain for integrable equations
which have a linear limit, suggests that  there  exist a $\bar{\partial}$ problem of a shape
\bge
\frac{{\partial}}{\partial \bar{\la}}\Phi (u,v,\la) = S_\gamma (u,v,\la) \Phi (u,v,\la),
\label{dbar}
\ene
where $S_\gamma (u,v,\la)$ is determined from (a yet to be formulated) boundary condition. 
We do not know how exactly to get this $\bar{\partial}$ problem here,
but since all examples known so far come in this shape, we conjecture it exist here as well. 
	Our Lagrangian is 
\bge
L = \dst\int_{\mbox{\gotic D}} \sqrt{ Det_{\al \be} \lb  \lt \dst\frac{\partial}{\partial u_\al}
f_{(0)}, \dst\frac{\partial}{\partial u_\be} f_{(0)} \rt   \rb } du_1 \wedge
du_2 = \dst\int_{\mbox{\gotic D}} \omega_{01} \wedge \omega_{02} 
\ene
with our choice of frame.
We assume we have a family of boundary conditions, for which we can resolve the (~\ref{dbar}) 
problem, and therefore we have a family of solutions of  (~\ref{dbar}) 
depending from the boundary condition $\Phi_{\gamma} (u,v, \la)$, which give rise to a family 
of one forms $\omega_{ij} = \bl (d \Phi_{\gamma} (u,v, 1) ) {\Phi_{\gamma}}^{-1} (u,v, 1)\br_{ij}$, 
depending from the boundary condition. We would like to compute in
 the second variation of the lagrangian with respect to change of boundary conditions, $\delta_1 \delta_2 L \equiv \delta_{f_o (t_1)} \delta_{f_o (t_2)}$ the term 
containing a delta function $\delta (t_1 -t_2)$. We will do it formally, assuming that there exist variation
$\delta$ commuting with the differential. 
Since 
$$
\bga
\delta \omega_{01} = \delta ( (d \Phi)  \Phi^{-1})_{01} =  
( (d \delta \Phi)  \Phi^{-1})_{01} - ( (d \Phi)  \Phi^{-1} \delta \Phi \Phi^{-1} )_{01} = \\
=( (d \delta \Phi)  \Phi^{-1})_{01} - \omega_{02} (  \delta \Phi \Phi^{-1} )_{21}; 
\ena
$$
here $\Phi\equiv \Phi(u,v,1) \in SO(1,n)$, and therefore the  symmetry conditions are \\
$(  \delta \Phi \Phi^{-1} )_{oo}=0, 
(  \delta \Phi \Phi^{-1} )_{o\al} = 
(  \delta \Phi \Phi^{-1} )_{\al o}, (  \delta \Phi \Phi^{-1} )_{\al \be} = -
(  \delta \Phi \Phi^{-1} )_{\be \al };  $ we used also our choice of the frame. We remark that
$\delta \Phi \Phi^{-1}$ are  zero forms on the tangent space, and $d \Phi \Phi^{-1}$ are one forms.

Then
$$
\delta (  \omega_{01} \wedge \omega_{02} ) = ( (d \delta \Phi)  \Phi^{-1})_{01} 
\wedge \omega_{02} +  \omega_{01} \wedge  ( (d \delta \Phi)  \Phi^{-1})_{02}.
$$
The only terms in the second variation which would contain a $\delta$ function would come
only from terms with the second derivative; as products of the first order derivatives cannot produce
a delta function. Therefore (terms with a delta function possible in the second variation ) are
$$
( (d \Delta \Phi)  \Phi^{-1})_{01} 
\wedge \omega_{02} +  \omega_{01} \wedge  ( (d \Delta  \Phi)  \Phi^{-1})_{02}
$$
Here $\Delta = \delta_1 \delta_2$
\vspace{5mm}

{\it Proposition} For a mean curvature zero surface $b_{(\mu),1} + b_{(\mu),2} =0, \mu =3,4,5,$ 
it follows from the 
Maurer Cartan equations, that the terms in the second variation which contain a $\delta$ function
depend only from the boundary condition, and given by
\vspace{10mm}

\bge
\dst\oint_{\partial \mbox{\gotic D}} {\lb( \Delta \Phi)  \Phi^{-1})\rb }_{01} \   \omega_{02}-  {\lb ( \Delta  \Phi)  \Phi^{-1} 
\rb}_{02} \  \omega_{01},
\ene
\vspace{10mm}

since 
\bge
\bga
( (d \Delta \Phi)  \Phi^{-1})_{01} 
\wedge \omega_{02} +  \omega_{01} \wedge  ( (d \Delta  \Phi)  \Phi^{-1})_{02} = \\[6mm]
=d \bl  {\lb( \Delta \Phi)  \Phi^{-1})\rb }_{01} \   \omega_{02}-  {\lb ( \Delta  \Phi)  \Phi^{-1} 
\rb}_{02} \  \omega_{01}  \br .
\label{loopisboundary}
\ena
\ene

Proof:
The difference between the right  hand side and the left hand side in 
(~\ref{loopisboundary}) 
is 
$$
\bga
{(( \Delta \Phi) \Phi^{-1})}_{o\mu} ( \omega_{1\mu} \wedge \omega_{02}  + \omega_{o1} \wedge \omega_{2\mu} ) + \\
{(( \Delta \Phi) \Phi^{-1})}_{o1} ( d \omega_{02} - \omega_{o1}\wedge \omega_{02} ) - \\
-{(( \Delta \Phi) \Phi^{-1})}_{o1} ( d \omega_{01} - \omega_{o2}\wedge \omega_{21} ) ;
\ena
$$
the first term is zero since  
$ ( \omega_{1\mu} \wedge \omega_{02}  + \omega_{o1} \wedge \omega_{2\mu} )=
 (b_{(\mu) ,1} + (b_{(\mu) ,1} )  \omega_{01} \wedge \omega_{02}$, and we have a 
$(b_{(\mu) ,1} + (b_{(\mu) ,1} )=0$ surface; the other terms are zero due to the 
Maurer Cartan, as it looks in our choice of the frame.

\newpage

\section{Acknowledgments}

Hospitality and financial support of  IHES and BRIMS institutes is appreciated.
I was introduced to some of the ideas and methods  
used here during a conversation with Prof. I.M. Gelfand and Dr. Juan Carlos
Alvarez Paiva, which occured at
IHES in summer, 1998. I am grateful to Prof. Gelfand and Prof. Fokas 
for discussions. I am also grateful to X. Y. for sticking my nose into 
\cite{hep-th/0002106} just a month after it appeared and then leaving me alone.

\end{document}